\documentclass{ws-p9-75x6-50}

\def\la{\mathrel{\hbox{\rlap{\hbox{\lower4pt\hbox{$\sim$}}}\hbox{$<$}}}}
\def\ga{\mathrel{\hbox{\rlap{\hbox{\lower4pt\hbox{$\sim$}}}\hbox{$>$}}}}

\def\ApJ{{\em Ap. J.}}

\def\AApS{{\em Astron. \& Ap. Supp.}}
\def\AAp{{\em Astron. \& Ap.}}
\def\BASI{{\em BASI}}
\def\MN{{\em MNRAS}}

\begin{document}

\title{Evidence For and Against Collimation of Gamma Ray Bursts}

\author{James E. Rhoads}

\address{Space Telescope Science Institute, 3700 San Martin Drive,
Baltimore, MD 21210, USA\\E-mail: rhoads@stsci.edu}


\maketitle

\abstracts{ The degree to which gamma ray bursts are collimated is now
the dominant uncertainty in their energy requirements and event rates.
In this review I begin with the reasons for studying GRB collimation,
then discuss existing tests for collimation and their applications to
date, and finally outline some possible future tests.  The most
important conclusions are that (1) mean collimation angles much
tighter than $1^\circ$ appear ruled out; (2) the collimation angle
appears to vary from burst to burst (like most other GRB properties).
Some alternative explanations of apparent collimation signatures
remain, but it should be possible to distinguish them from true
collimation with future data sets and may be possible already.  New
satellites, improved followup observations, and new tests for
collimation all promise continued rapid progress in coming years.}

%

\section{Introduction}
There are two main reasons to look for evidence of collimation in
gamma ray bursts (GRBs).  First, collimation is an extremely common
phenomenon in astrophysical outflows.  Jets have been observed on
scales ranging from parsecs (protostellar objects) up to tens of
kiloparsecs (radio galaxies) and with speeds ranging from $v \sim 100
\hbox{km}/\hbox{s}$ (protostellar objects) up to $\Gamma \sim 10$
(active galactic nucleus jets).  It is thus natural to suspect that
GRBs may also be collimated.


Second, the energy requirements of gamma ray bursts have long been
cause for discussion, and in some quarters concern.  Present data show
these requirements to exceed $10^{54}$ ergs in gamma ray production
for the most energetic bursts {\it if} the bursts are isotropic.  Such
energies are regarded as an ``energy crisis'' for some classes of
burst progenitor models.  Collimation allows these energy requirements
to be relaxed by a factor of $\Omega_\gamma / (4 \pi)$, where
$\Omega_\gamma$ is the solid angle into which the gamma rays are
emitted.  At the same time, the required event rate of GRBs scales as
$4 \pi / \Omega_\gamma$.  Constraining $\Omega_\gamma$ is thus a
necessary prerequisite for knowing what classes of object can produce
the observed gamma rays and how common or rare burst progenitors must
be.

By considering the relativistic nature of GRBs, we can see both the
reasons collimation angles were unconstrained in past and the means
for current progress in the field.  Until 1997, all of our information
on GRBs came from gamma ray observations.  These observations showed
that (a) GRBs are highly energetic, with fluences $\sim 10^{-5}$ erg
cm$^{-2}$ ``representative'' for bright bursts; (b) burst light curves
are highly structured with variability on time scales down to $\sim
10^{-3}$ s; and (c) bursts have nonthermal spectra reasonably
described by broken power laws.  These three characteristics imply
that (a) burst energies are $10^{52}$ erg and above for cosmological
distances and isotropic radiation; (b) the bursters are small, with a
characteristic size na\"{\i}vely estimated at $10^{-3} \hbox{s} \times
c \approx 3 \times 10^7 \hbox{cm}$; and (c) the gamma rays must come
from an optically thin region to avoid being thermalized.  However,
the energy density implied by (a) and (b) would require a very high
density of gamma rays in a static source model, with a consequent
optical depth $\tau_{\gamma \gamma} \gg 1$ for $\gamma + \gamma
\rightarrow e^- + e^+$ and an expectation of a thermal emergent
spectrum.  To resolve this apparent conflict, it is sufficient and
probably necessary for the source to be in bulk relativistic
expansion, which relaxes both the constraint on source size (since
observed variations appear more rapid when the emitting matter is
approaching at nearly lightspeed) and the pair creation optical depth
(since the photons in question can have energies much below $511
\hbox{keV}$ in a relativistically moving frame).
(See Paczy\'{n}ski 1986; Goodman 1986; Krolik \& Pier 1991).  The
implied minimum Lorentz factors are in the range $100 \la \Gamma \la 1000$
(Woods \& Loeb 1994), making GRBs the best known astrophysical
laboratory for the study of ultrarelativistic shocks.

These high Lorentz factors immediately imply the possibility of
substantial collimation.  Burst ejecta that emit isotropically in their
rest and move with Lorentz factor $\Gamma$ relative to the lab frame
will appear to beam their emission forward into a cone of
characteristic opening half angle $1/\Gamma$.  This implies that
whether or not the bursts are isotropic, the gamma rays that we
observe come from material moving within angle $1/\Gamma \la 1/100
\; \hbox{radian}$ of the line of sight, and we cannot determine the
presence or absence of ejecta outside that narrow cone from these
gamma rays alone.  The reduction in energy requirements from
collimation could therefore be as extreme as $10^{-5} \ga
\Omega_\gamma / (4 \pi) \ga 10^{-7}$ for $100 \la \Gamma \la 1000$.

The fast ejecta from GRBs must eventually encounter a sufficient
column density of ambient medium to slow down.  A natural consequence
is that the bulk kinetic energy of the ejecta is converted into other
forms, and some can be radiated in an afterglow (Paczy\'{n}ski \&
Rhoads 1993; Katz 1994; M\'{e}sz\'{a}ros \& Rees 1997).  Because the
afterglow allows us to probe the evolution of the GRB remnant while
the Lorentz factor decreases from $\Gamma_0$ to $\sim 1$, afterglow
studies open the way for comparatively straightforward tests of GRB
collimation.

\section{Orphan Afterglows}
As the GRB blast wave sweeps up the ambient medium and decelerates,
the peak emission frequency of the afterglow becomes lower (scaling as
$\Gamma^4$ in the simplest models) while the degree of relativistic
beaming lessens ($\Omega_\gamma \propto \Gamma^{-2}$).  Thus,
observers situated too far from the axis of the ejecta to see the GRB
itself may begin to see the afterglow at some point in its evolution,
and the observed event rate increases with wavelength.  The statistics
of such ``orphan afterglows'' can be used to constrain the mean
collimation angle of the GRB population (Rhoads 1997; Perna \& Loeb
1998).  Finite detection thresholds introduce some subtlety in
applying this method: Limits can be placed on the ratio of opening
angles at two observed wavelengths in a model-independent way, but
actual measurements of the opening angle ratio require some knowledge
of the flux ratio distribution at these wavelengths.

This method has now been applied in at least a preliminary way to
X-ray, optical, and radio wavelengths.  The present state of the art
in X-ray orphan afterglows is work of Greiner (1999), who searched the
ROSAT All-Sky Survey data for transient events, finding a sample
of 23 sources.  However, their optical followup of the X-ray events
showed that red stars were present in essentially all cases, and
that the subset of these stars observed spectroscopically all had dMe
star spectra.  The observed transients were therefore mostly or
entirely stellar flares.  After accounting for this foreground,
Greiner et al conclude that no difference in collimation between the
$\gamma$-ray and X-ray events is required by the data.  Similar conclusions
were reached by Grindlay (1999) based on Ariel 5 data.

At optical wavelengths, many existing data sets may be suitable for
orphan afterglow tests (e.g., high redshift supernova searches,
microlensing studies, or any other large variability survey with time
sampling of a few days).  One particularly suitable data set where I
know that an orphan afterglow analysis is well underway is work of
Schaefer et al (2001), who have surveyed an area of $\sim 200$ square
degrees repeatedly to depths $R \sim 21$ over the course of $\sim 2$
weeks.  Present analysis shows no viable orphan afterglow candidates
in their data (Schaefer 2000, private communication).  This data set
would be expected to contain $\sim 0.2$ afterglows if bursts are
isotropic, so the absence of orphan afterglows suggests $\Omega_{{\rm
opt}} / \Omega_{\gamma} \ll 100$, which is enough to rule out the most
extreme collimation scenarios.

Finally, at radio wavelengths, Perna \& Loeb (1998) have used
published source counts and variability studies to place a limit on
the collimation angle, $\theta_\gamma \ga 5^\circ$.  (Because radio
afterglows last into the nonrelativistic phase of the GRB remnant
evolution, the radio afterglows are expected to radiate essentially
isotropically, and the orphan afterglow limits on $\Omega_{\rm radio}
/ \Omega_\gamma$ immediately imply a limit on $\Omega_\gamma$ itself.)
The uncertainty in this calculation is difficult to assess, because
Perna \& Loeb combined three different radio surveys at two different
wavelengths.

The main worry in applying orphan afterglow tests is that other
classes of transient source may provide a false positive signal.
Greiner et al have already demonstrated the importance of identifying
such foreground sources at X-ray wavelengths.  In general, both the
light curve and spectral energy distribution of a transient source
should match expectations for collimated afterglows if the source is
to be a viable orphan afterglow candidate.  A good screening tool for
possible candidates is that the spectral energy distribution should
usually be consistent with a power law, implying a specific
color-space locus for afterglows (Rhoads 2001).  Another tool is the
spatial distribution of orphan afterglows, which should presumably be
isotropic like that of GRBs and unlike many classes of foreground
variable source.  The most subtle class of confusing source may turn
out to be ``dirty fireballs'': Events like GRBs where the initial
ejecta suffer from more baryon pollution and hence have initial
Lorentz factors $1 \ll \Gamma_0 \ll 100$.  Such events will not
produce GRBs (for which $\Gamma_0 \ga 100$) but can certainly produce
afterglows.  Fortunately, the light curve method (below) should serve to
distinguish isotropic dirty fireballs from orphan afterglows of
collimated GRBs.

\section{Light Curve Breaks}
The most widely applied tests for GRB collimation so far derive
from light curve breaks expected in the afterglows of collimated
bursts (Rhoads 1997b, 1999; Sari, Piran, \& Halpern 1999).
Conceptually, there are two such breaks:  The first occurs when the
angle for relativistic beaming $1/\Gamma$ exceeds the collimation
angle $\zeta_m$ of matter in the jet, and the second occurs when
sideways expansion of the jet material increases the working surface
of the expanding remnant enough to affect its dynamics.  While the
distinction between these breaks has been emphasized by some (e.g.,
Panaitescu \& M\'{e}sz\'{a}ros 1999), they may be difficult to
distinguish observationally because the
time between them is less than the characteristic time for either
break to occur.  These breaks should be achromatic, i.e., should occur
at the same time at all wavelengths.

The existence of such breaks implies two tests for collimation.
First, good light curves (at any wavelength) can be used to detect the
break directly.  Second, the expected relation between spectral slope
and light curve slope differs before and after the break, so that by
measuring both it may be possible to identify collimated bursts even
when the break preceded the first followup observations.  Light curve
breaks have now been directly observed in GRBs 990123 (Castro-Tirado et
al 1999; Fruchter et al 1999; Galama et al 1999; Kulkarni et al 1999),
990510 (Stanek et al 1999; Harrison et al 1999; Kumar \& Panaitescu 2000a),
000301C (e.g. Berger et al 2000; Masetti et al 2000; Sagar et al 2000b;
Rhoads \& Fruchter 2001), 000926 (Sagar et al 2000c), and (weakly)
991216 (Halpern et al 2000), while the absence of such breaks has been
shown for GRB 970508 (Rhoads 1999b).  The inferred opening angles
range from $\sim 2.5^\circ$ (GRB 000301C, Rhoads \&
Fruchter 2001, if the break is indeed due to collimation) to lower
limits $> 30^\circ$ (GRB 970508, Rhoads 1999b).  Comparisons of
spectral and light curve slopes have lead to inferences of collimation
for several more bursts, including GRBs 980519 (Halpern et al 1999;
Sari, Piran, \& Halpern 1999), and 991208 (Sagar et al 2000a).

These data remain open to alternative interpretations, however.
Spectral and light curve slopes for isotropic bursts expanding into a
stellar wind ambient medium (with density $\rho \propto r^{-2}$) can
closely resemble the slopes expected from the post-break regime of a
collimated afterglow (Chevalier \& Li 1999; Halpern et al 1999).
Also, the transition to the nonrelativistic regime can introduce a
break into the afterglow light curve that may account for some of the
observed breaks (Dai \& Lu 1999).  A third class of break can occur when
the ambient medium has a sharp drop in density beyond some radius.
Kumar \& Panaitescu (2000b) explored this model in detail and find that a
light curve slope as steep as $t^{-3}$ can result, potentially
explaining the behavior of GRB 000301C.  Still, these other classes of
break should produce a different relation between spectral and light
curve slope that would be measurable given an adequate multiwavelength
monitoring campaign.

Another unresolved issue is the sharpness of the observed breaks.  The
transition between the two asymptotic light curve slopes is expected
to be rather slow in the models (Rhoads 1999a; Moderski, Sikora, \&
Bulik 2000; Panaitescu \& Meszaros 1999).  The most detailed analysis
of this problem so far is by Kumar \& Panaitescu (2000a) who found
that the GRB 990510 break can be reproduced by combining a collimation
break with the light curve break that occurs when the peak of $f_\nu$
or $\nu f_\nu$ shifts through the observed bandpass.  This requires a
coincidence that is acceptable once or a few times, but if such
fortuitous circumstances appear in a large sample of afterglows it
will indicate a problem with the model.

\section{Polarization}
GRB afterglows are thought to arise from synchrotron radiation in the
expanding burst remnant.  It is thus natural to expect some degree of
polarization.  However, if the burst is spherically symmetric, it is
likely that the {\it net\/} polarization of the afterglow will be
small, since there is no preferred position angle.  Jetlike bursts allow
the symmetry to be broken:  If we are not precisely on the jet axis,
then the projection of the jet axis on the sky defines a preferred
direction and the polarization ought to be either parallel or
perpendicular to that direction.  Ghisellini \& Lazatti (1999) and
Sari (1999) have modeled the polarization of collimated burst
afterglows and find that multiple polarization peaks are expected
(at least two peaks, and three if sideways expansion of the ejecta is
substantial).  The position angle changes by $90^\circ$ between peaks.
Observationally, this test has yet to be applied in a meaningful way,
though the technology exists to do so.  Several groups have reported
attempting polarization measurements, resulting in one published
measurement and a few upper limits.  However, to really see multiple
peaks would require $\ga 10$ epochs of polarization measurement, far
beyond anything yet achieved.

\section{GRB Remnant Statistics}
Because collimation determines both GRB rates and energies, the
numbers and sizes of GRB remnants are expected to scale with the
collimation angle.  Thus, if we can identify GRB remnants in a
reasonably complete and unambiguous way, we can infer the event rate
from their statistics.  HI supershells were proposed as possible GRB
remnants in 1998 (Efremov, Elmegreen, \& Hodge 1998; Loeb \&
Perna 1998).  However, there are more conventional explanations for
supershells (OB association winds and/or multiple supernovae) that can
produce energies comparable to an isotropic GRB and must be ruled out
before any shell would be a compelling GRB remnant candidate.  The
best test so far suggested (Perna, Raymond, \& Loeb 2000) is to look
for recombination lines of the highly ionized species that will be
produced by GRBs (with their hard UV and X-ray photons) but not by OB
stars or supernovae.  Like the polarization test, this method relies
on reasonably proven technology but has yet to bear fruit.

It may also be possible to study collimation through GRB
remnant shapes.  A collimated burst will yield a long, narrow remnant
for a time.  However, this test does not seem immediately promising,
both because of the high angular resolution required and because the
remnant will likely become fairly round by the time the blast wave becomes
nonrelativistic.  A further difficulty is that for collimated bursts
that we see, we will be nearly on-axis and will therefore see a fairly
round projected image of the remnant.

\section{Further Possibilities}
The tests discussed above all rely (directly or indirectly) on the
electromagnetic radiation from GRBs and their afterglows.  In future,
it may become possible exploit other types of astronomy for GRB
collimation.  In particular, coalescing neutron star binaries are a
promising class of GRB progenitor models, and are expected to be among
the more readily detected gravitational wave sources in the universe.
The angular distribution of these gravitational waves should be
calculable.  Moreover, the ``radiative transfer'' problem for these
gravitational waves is simpler than that for the gamma rays and lower
energy afterglow photons, for which many types of scattering and
absorption potentially affect the observed properties of the burst.
If neutron star inspiral events do indeed lead to GRBs, then, it
should be possible to constrain the collimation of the gamma rays by
comparing statistics of GRBs to statistics of gravitational wave
bursts.  Similar arguments may be applied to neutrino emission from
GRBs, though the angular distribution of the neutrinos will probably
depend more heavily on details of the model than would gravitational
waves.

\section{Conclusions}
The study of gamma ray burst collimation is still a very young field,
and though there is much yet to learn we have come a long way from
what we knew at the beginning of 1997.  This progress has been fueled
in two ways by by the discovery of afterglows.  First, by yielding GRB
redshifts, afterglows finished the GRB distance scale debate.  This
left collimation as the dominant uncertainty in burst energy
requirements and event rates, motivating research on collimation.
Second, afterglows have provided most of the tools for constraining
collimation angles discussed above, including both tools that have
found substantial applications to date.

Two primary conclusions emerge from the data so far.  First, the most
extreme collimation scenarios (those with opening angles $\sim
1/\Gamma_0 \la 1/100$ radian) appear to be ruled out by nondetections
in orphan afterglow searches.  The mean opening angle favored by these
searches (which is only a lower bound) is a factor of 10 larger, $\ga
5^\circ$.  Second, the collimation angle appears to vary substantially
from burst to burst, based on the observed light curve breaks.  While
this statement might still be challenged (e.g. if some observed breaks
are not due to collimation), it is consistent with the huge variations
seen in other GRB properties (e.g., light curve shapes).
Interestingly, the inferred collimation angle and the isotropic equivalent
energy appear to anticorrelate, so that GRBs are more nearly standard
candles after collimation corrections are applied (see Frail et al 2001
for further discussion of this possibility).  

The current rapid progress in studying GRB collimation should continue
in the coming decade.  New high energy missions will provide better
samples of bursts.  Coordinated multiwavelength followup of these
bursts will increase the sample of bursts with breaks (especially
those with breaks at early times), and will yield data sets that can
more easily distinguish between different classes of light curve
break.  The first good polarization curves for GRB afterglows are
another likely result of increased rates of prompt, accurate burst
locations.  Finally, large variability surveys are becoming much easier
thanks to new instruments, and will soon either find orphan afterglows or
yield truly compelling limits on their frequency.  And once we have
measured the collimation of gamma ray bursts, we can begin to use
their energy requirements and event rates as real constraints on
progenitor models for the first time.

\end{document}